\documentclass[preprint2]{aastex}

\shorttitle{Empirical models for Dark Matter Halos. II.}
\shortauthors{Graham et al.}
\begin{document}

\title{Empirical Models for Dark Matter Halos. II. Inner profile slopes, 
dynamical profiles, and $\rho/\sigma^3$}

\author{Alister W.\ Graham\altaffilmark{1}}
\affil{Mount Stromlo and Siding Spring Observatories, Australian National 
University, Private Bag, Weston Creek PO, ACT 2611, Australia.}
\altaffiltext{1}{Graham@mso.anu.edu.au}

\author{David Merritt}
\affil{Department of Physics, Rochester Institute of Technology, 
Rochester, NY 14623, USA.}

\author{Ben Moore}
\affil{University of Zurich, Winterthurerstrasse 190, CH-8057, 
Z\"urich, Switzerland.}

\author{J\"urg Diemand}
\affil{Department of Astronomy and Astrophysics, University 
of California, 1156 High Street, Santa Cruz, CA 95064, USA.}

\author{Bal{\v s}a Terzi\'c}
\affil{Department of Physics, Northern Illinois University, 
DeKalb, IL 60115, USA.}

\begin{abstract}

We have recently shown that both the Prugniel-Simien model and
S\'ersic's function (hereafter referred to as the Einasto model when
applied to internal density profiles) describe simulated dark matter
halos better than an NFW-like model with equal number of parameters.
Here we provide analytical expressions for the logarithmic slopes of
these models, and compare them with data from real galaxies.
Depending on the Einasto parameters of the dark matter halo, one can
expect an extrapolated, inner (0.01--1 kpc), logarithmic profile slope
ranging from $\sim -0.2$ to $\sim -1.5$, with a typical value at 0.1
kpc around $-0.7$.  Application of this (better fitting) model
therefore alleviates some of the past disagreement with observations
on this issue.
We additionally provide useful expressions for the concentration and
assorted scale radii: $r_s, r_{-2}, r_{\rm e}, R_{\rm e}, r_{\rm
vir}$, and $r_{\rm max}$ --- the radius where the circular velocity
profile has its maximum value.  We also present the circular velocity
profiles and the radial behavior of $\rho(r)/\sigma(r)^3$ for both the
Einasto and Prugniel-Simien models, where $\sigma(r)$ is the velocity
dispersion associated with the density profile $\rho(r)$.  We find
this representation of the phase-space density profile to be well
approximated by a power-law with slope slightly shallower than $-2$
near $r=r_{-2}$.

\end{abstract}

\keywords{
dark matter ---
galaxies: fundamental parameters --- 
% galaxies: clusters: general ---
galaxies: halos
galaxies: structure --- 
methods: analytical 
}

\section{Introduction}

In an interesting turn of events, S\'ersic's (1963, 1968) 3-parameter
function, developed to describe the projected (on the plane of the
sky) radial stellar distributions of galaxies, has been shown to also
match the internal (3D) density profiles of simulated dark matter halos
(Navarro et al.\ 2004).  Intriguingly, this function was shown to
provide a better fit than the 3-parameter NFW-like model with
arbitrary inner power-law slope (Diemand, Moore, \& Stadel 2004;
Merritt et al.\ 2005, 2006).  The functional form of S\'ersic's
equation was independently developed by Einasto (1965, 1968, 1969) and
used to describe the internal density profiles of galaxies (see also
Einasto \& Haud 1989 and Tenjes, Haud, \& Einasto 1994).  We shall
therefore refer to this function as the S\'ersic model when applied to
projected distributions, and as the Einasto model when applied to
internal density profiles.

One of the main concerns in (Merritt et al.\ 2005, 2006) was whether
the deprojected form of S\'ersic's function might provide a better
description of the density profiles than the Einasto model.
Specifically, in Merritt et al.\ (2006, hereafter Paper I), the
analytical approximation to the deprojection of S\'ersic's function
given in Prugniel \& Simien (1997) was tested along with the Einasto
model, the NFW-like model, and various other fitting functions.
Overall, the Prugniel-Simien and Einasto models performed the best,
providing a good description to both the galaxy- and cluster-sized
halos built from hierarchical $N$-body simulations, and also the monolithic
cold collapse halos.\footnote{In passing we note that the 3-parameter
anisotropic Dehnen-McLaughlin (2005) model matched the galaxy-sized
halos best, but it did not perform so well in describing the
cluster-sized halos, and it was unable to describe the halos formed from
spherical cold collapses.}  Curiously, while the Prugniel-Simien
model provided the best fit to the cluster-sized halos, Einasto's
model provided a better fit to the galaxy-sized halos.  In this paper
we explore some of the properties of these two models and some of the
consequences they imply, and comparisons they enable, with real
galaxies and galaxy clusters.

In an effort to help clarify and unify the various parameters of the
different models, Section~2 provides relations between such quantities
as effective radius $R_{\rm e}$, virial radius $r_{\rm vir}$, 
the radius where the logarithmic slope of the model equals $-2$, $r_{-2}$,
and `concentration' as measured by observers and by modelers. 

In Section~\ref{SecCube} we present the phase-space density 
profiles, or more specifically, the density profiles divided by the
cube of their associated velocity dispersion profiles, showing how, for
sufficiently large shape parameters $n$, both the Einasto
model and the Prugniel-Simien model approximate a power-law 
$\sim r^{-2}$ near $r_{-2}$.  

In Section~\ref{SecSlope} we derive the logarithmic slopes of the
Einasto and Prugniel-Simien model, and compare these with real data.
This is of particular interest because the innermost slope of these
models is considerably shallower than $-1$ and in fact equal to zero
in the case of the Einasto model at $r=0$.  The inward extrapolation
of these models, rather than the NFW-like model, therefore noticeably
reduces the disagreement between modelers and observers on this issue.

Our findings are summarized in Section~\ref{SecSum}.

\section{The models: assorted radial scales and concentration}\label{Defin}

We will discuss three (3-parameter) empirical models used for
describing the internal density profiles of galaxies, clusters, and
halos.
Each model has three parameters and their application to simulated
halos can be seen in Paper I.

The first model is an adaptation of the Navarro, Frenk, \& White (1995,
hereafter NFW) model to give a double power-law with an outer slope of
$-3$ and an arbitrary inner slope denoted by $\gamma$.  The radial
density profile, $\rho(r)$, of this model can be written as
\begin{equation}
\rho(r) = \frac{2^{3-\gamma}\rho_s}{(r/r_s)^{\gamma}(1+r/r_s)^{3-\gamma}}, 
\label{EqNFW}
\end{equation}
where $r_s$ is the scale radius at the density $\rho_s$, marking the
center of the transition between the inner and outer power-laws with
(extrapolated) slopes of $-\gamma$ and $-3$.  This function represents
a restricted form of the more generic ($\alpha, \beta, \gamma$) model
(Hernquist 1990, his equation~43; see also Zhao 1996) 
and we shall refer to it as the ($1, 3, \gamma$) model.
Setting ($\alpha, \beta, \gamma$)=(1, 3, 1) yields the NFW model,
while (1.5, 3, 1.5) gives the model in Moore et al.\ (1999).
Because the ($1, 3, \gamma$) model has already been studied in detail,
our main focus will be on the models of Einasto (1965) and Prugniel \&
Simien (1997).  

Einasto's model is given by the equation
\begin{equation}
\rho(r)=\rho_{\rm e} \exp\left\{ -d_n\left[ (r/r_{\rm e})^{1/n} -1\right] \right\}. 
\label{SerDen}
\end{equation}
The parameter $n$ describes the shape of the density profile.  Larger
values of $n$ result in steeper inner profiles and shallower outer
profiles.  The quantity $d_n$ is defined to be a function of $n$ such
that $\rho_{\rm e}$ is the density at the {\it effective} radius $r_{\rm e}$ which
encloses a volume containing half of the total mass.  A good
approximation for $n \gtrsim 0.5$ is given by $d_n \approx 3n -
1/3 + 0.0079/n$ (Mamon 2005, priv comm.), although we have used the
exact value coming from $\Gamma (3n)=2\times\gamma (3n,d_n)$, where
$\gamma(x_1,x_2)$ and $\Gamma(x)$ are the incomplete and complete gamma
functions, respectively (see Paper I).  
The mass profile (Mamon \& {\L}okas 2005, their equation~A2; 
Cardone et al.\ 2005, their equation~11) is given by 
\begin{equation}
M(r) = 4\pi n r_{\rm e}^{3} \rho_{\rm e} {\rm e}^{d_n} {d_n}^{-3n} \gamma (3n,x). 
\label{Sermass}
\end{equation}
We shall at times refer to the value of $n$ from Einasto's model
as $n_{\rm Ein}$.

The Prugniel-Simien model can be expressed as 
\begin{equation}
  \rho(r) = \rho^{\prime} \left({r\over R_{\rm e}}\right)^{-p}
             \exp\left[-b_n\left( r/R_{\rm e} \right)^{1/n} \right],
\label{EqPS97}
\end{equation}
with
\begin{equation} \label{projIe}
  \rho^{\prime} = {M\over L} \ I_{\rm e} {\rm e}^{b_n} \ {b_n}^{n(1-p)} \ {\Gamma(2n) \over
            2 R_{\rm e} \Gamma(n(3-p)) }.
\end{equation}
Once again, the parameter $n$ describes the curvature of the density 
profile.  The quantity $b_n$ is a function of $n$ defined in such a way 
that $R_{\rm e}$ is the {\it effective} radius containing half of the 
total mass when the 3D
sphere defined by this density profile is seen in projection onto a 2D
plane.  Although we use the exact solution for $b_n$, coming from
$\Gamma (2n)=2\times\gamma (2n,b_n)$ (see Graham \& Driver 2005),
$b_n$ can be approximated by $2n-1/3+0.009876/n$ for values of $n
\gtrsim 0.5$ (Prugniel \& Simien 1997).
In addition to $n$ and $R_{\rm e}$, the third parameter which one
solves for, $\rho^{\prime}$, is defined so that the volume-integrated
mass from equation~\ref{EqPS97} is equal to the area-integrated mass
of a S\'ersic function with the standard parameter set $I_{\rm e},
R_{\rm e},$ and $n$ (see Paper I).

The final term $p$ is not a parameter but instead, like $b_n$, another
function of $n$.  It is chosen to maximize the agreement between the
Prugniel-Simien model and the deprojected S\'ersic model having the
same parameters $n$ and $R_{\rm e}$. A good match is obtained when
$p = 1.0 - 0.6097/n + 0.05463/n^2$ (Lima Neto et al.\ 1999; 
see also Paper I, their figure 13). 
The value of $p$ is also responsible for determining the logarithmic
profile slope at small radii.   Setting $p$ to zero, the Prugniel-Simien
model has the same functional form as Einasto's model. 

The internal
density of the Prugniel-Simien model 
at $r=R_{\rm e}$ is given by $\rho_{\rm e}=\rho^{\prime} {\rm
e}^{-{b_n}}$, and the projected surface density at $R=R_{\rm e}$,
denoted by $I_{\rm e}$, can be solved for using equation~\ref{projIe}. 
Comparisons of dark matter halos (fitted with the
Prugniel-Simien model) with real galaxies (fitted with S\'ersic's
model) is obviously remarkably easy using this model. 
The associated mass profile 
(Lima Neto et al.\ 1999; M\'arquez et al.\ 2001) is given by the equation
\begin{equation} \label{PSmass}
M(r)  = {4\pi n {R_{\rm e}}^3} \rho^{\prime} {{b_n}^{-(3-p)n}}
\gamma\left((3-p)n,Z\right),
\end{equation}
where $Z \equiv b_n(r/R_{\rm e})^{1/n}$. 
Expressions for the associated gravitational potential, force, 
and velocity dispersion can be found in Terzi\'c \& Graham 2005). 
In what follows, we shall at times refer to the value of $n$ from the 
Prugniel-Simien model as $n_{\rm PS}$.  

In the following subsections we explore a number of important radii
associated with the above models, and address the issue of `concentration'.

\subsection{The peak in the $4\pi G r^2 \rho(r)$ 
{\rm [km s$^{-1}$]$^2$} profile, $r_{-2}$}

As will be discussed in Section~\ref{SecSlope}, the scale radii $r_{\rm e}$
of the Einasto model, and the (projected) half-mass radii $R_{\rm
e}$ from the Prugniel-Simien model, occur where the logarithmic slope
of the density profile is $\sim -3$.  This can be quite far out, 
and so we define an additional radial scale. 
We do so by obtaining the radius where the profile $4\pi G r^2
\rho(r)$, which has units of velocity squared, has its maximum.  The
integral of this profile gives the enclosed mass.

For the (1, 3, $\gamma$) model (equation~\ref{EqNFW}), this maximum
occurs at a radius that we denote by $r_{-2,(1, 3, \gamma)}$, such that 
\begin{equation} \label{NFWR2}
r_{-2,(1, 3, \gamma)} = (2-\gamma)r_s,   \hskip30pt  \gamma < 2. 
\end{equation}
When $\gamma = 1$, as in the NFW model, the radius $r_{-2,(1, 3, \gamma)} =
r_s$, and when $\gamma = 1.5$ one has $r_{-2,(1, 3, \gamma)} = r_s/2$.  It turns
out that the radius $r_{-2,(1, 3, \gamma)}$ corresponds to the point where the
logarithmic slope of the (1, 3, $\gamma$) density profile equals $-2$,
hence the adopted nomenclature.  Similarly for the Einasto and
Prugniel-Simien density model, solving where the derivative of the
profile $4\pi G r^2 \rho(r)$ equals zero, one finds that these
profiles also peak at the radius where their logarithmic slope equals $-2$.
This is easy to understand when one notes that the solution to 
${\rm d}[r^2 \rho(r)]/{\rm d}r = 0$ leads to $d \rho/\rho = -2 dr/r$. 
It is also easy to show that this corresponds to a maximum 
for any density profile with a monotonically decreasing slope. 
% 
%  with alpha = d ln rho / d ln r,
%        d^2(r^2 rho)/dr^2 = rho (r dalpha/dr + alpha^2 + 3 alpha + 2)
%  and with alpha=-2,
%        d^2(r^2 rho)/dr^2 = rho r d alpha/dr
%  hence
%        sgn(d^2(r^2 rho)/dr^2 = sgn[d(d ln rho/d ln r)/dr]. 
%  i.e. if alpha is becoming more negative (steeper slope, as in LCDM
%  halos), the 2nd derivative of r^2 rho is negative, hence the radius of
%  slope -2 is indeed a maximum of r^2 rho.

For Einasto's model (equation~\ref{SerDen}), one has 
\begin{equation} \label{SerR2}
r_{-2,{\rm Ein}} = \left( \frac{2n_{\rm Ein}}{d_n} \right)^{n_{\rm Ein}} r_{\rm e}, 
\end{equation}
and for the Prugniel \& Simien density profile (equation~\ref{EqPS97})
one has 
\begin{equation}  \label{PSR2}
r_{-2,{\rm PS}} = \left( \frac{n_{\rm PS}(2-p)}{b} \right)^{n_{\rm PS}} R_{\rm e}. 
\end{equation}
When $n_{\rm Ein}=6$, $r_{-2,{\rm Ein}} \sim 0.10 r_{\rm e}$.
When $n_{\rm PS}=3$, $r_{-2,{\rm PS}} \sim 0.25 R_{\rm e}$.
(These representative values of the shape parameter have been 
taken from Paper~I, which applied the above models to a number
of simulated dark matter halos.) 

% From equations~\ref{SerGam} and \ref{PSGam}, one can easily verify
% that the radius $r_{-2}$ corresponds to point where the logarithmic
% slope of the Einasto and Prugniel-Simien density profiles equals $-2$.

% xxx:  What is the ratio of $4\pi G r^2 \rho(r_{-2}) / {\sigma_{max}}^2$
% for different n??  Is it (nearly) constant?  That is, can we use the
% density profile to predict the maximum vel.disp?  note: they peak at
% different radii, is this offset in where they peak correlated with the
% above ratio (if it isn't constant)? 

Evaluating Einasto's model (equation~\ref{SerDen}) at $r_{-2,{\rm Ein}}$ to
give the density $\rho_{-2,{\rm Ein}}$, and expressing $r_{\rm e}$ in terms
of $r_{-2,{\rm Ein}}$ (equation~\ref{SerR2}), Einasto's model can be written 
as\footnote{For clarity, we have dropped the subscript `Ein' from the 
parameter $n$.} 
\begin{equation} \label{SerNav}
\rho_{\rm Ein}(r) = (\rho_{-2,{\rm Ein}}) \times
\exp\{-2n[(r/r_{-2,{\rm Ein}})^{1/n} -1]\}, 
\end{equation}
where 
\begin{equation}
\rho_{-2,{\rm Ein}} = \rho_{\rm e}{\rm e}^{d-2n}. 
\end{equation}
This is the expression used in Navarro et al.\ (2004). 

% Using this form for the Einasto profile does not affect the
% best-fitting n-value (nor the rms residual). It gives r_{-2}
% rather than r_e, and thus the density at this radius is of course
% different to that at r_e. 

Re-expressing Prugniel \& Simien's model in terms of $r_{-2}$, one 
has\footnote{For clarity, we have dropped the subscript `PS' from the 
parameter $n$.}
\begin{eqnarray} \label{PSNav}
\rho_{\rm PS}(r) & = & (\rho_{-2,{\rm PS}}) \left( \frac{r}{r_{-2,{\rm PS}}} \right)^{-p} \nonumber \\
   & & \hskip-50pt  \times  \exp\{-n(2-p)[(r/r_{-2,{\rm PS}})^{1/n} -1]\}, 
\end{eqnarray}
where 
\begin{equation}
\rho_{-2,{\rm PS}} = \rho^{\prime} \left( \frac{b}{n(2-p)} \right)^{np} 
\exp\{ n(p-2)\}.
\end{equation}

% Juerg wrote: {\sl The scale radius (I mean where the slope reaches -2)
% is an interesting physical scale, it is related to the formation time
% of a halo and after that it is roughly constant in physical units,
% except during major mergers.}

\subsection{The peak in the circular velocity profile, $r_{\rm max}$} 

\begin{figure}
\includegraphics[scale=0.44,angle=270]{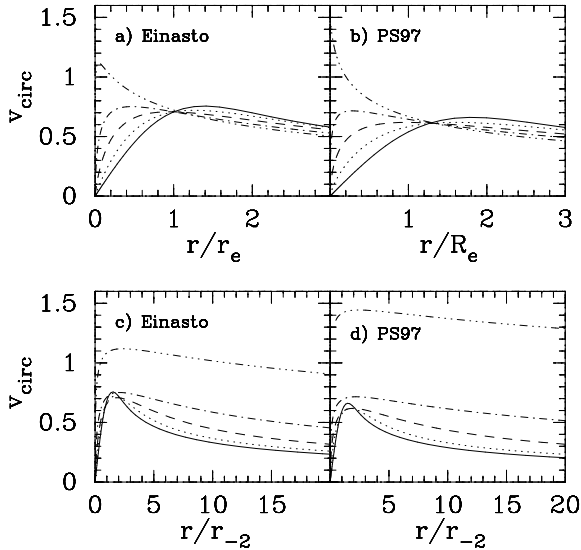}
\caption{
Circular velocity profiles for 
a) the Einasto model (equation~\ref{SerDen}), and b) Prugniel
\& Simien's model (equation~\ref{EqPS97}), 
for varying values of the profile shape 
$n$: $n=0.5$ (solid lines), $n=1$ (dotted), $n=2$ (dashed),
$n=4$ (dash-dot), $n=10$ (dash-triple dot).
The lower panels show the same thing except the radius has
been normalized against $r_{-2}$ (the radius where the 
logarithmic slope of the density profile equals $-2$) rather 
than the effective radii of the models.  Panel c) Einasto's model, 
panel d) Prugniel \& Simien's model. 
$n=0.5$ (solid line); $n=1$ (dotted line); $n=2$ (dashed line); 
$n=4$ (dot-dash line); $n=10$ (triple-dot-dash line). 
}
\label{FigVel}
\end{figure}

While an isothermal density profile, $\rho(r) \propto r^{-2}$, has a
flat rotation curve, the radius where the logarithmic slope of our
(non-isothermal) 
density profiles equals $-2$, $r_{-2}$, does not coincide with the
flat portion of the rotation curve, i.e. where the rotation curve has
its maximum value.
The rotation curves are simply given by the circular velocity profiles: 
${v_{\rm circ}}^2 (=GM(r)/r)$, with $M(r)$ defined previously. 
The maximum occurs at a radius $r_{\rm max}$ which is larger than
$r_{-2}$, and is shown in Figure~\ref{FigVel} as a function of 
both the effective radius and $r_{-2}$ for the Einasto and 
Prugniel-Simien models. 
% (see also Figure~11 in Terzi\'c \& Graham 2005 for the Prugniel \&
% Simien density profile, and Figure~3 in Cardone et al.\ (2005) for
% Einasto's model.).

The width near the peak of the circular velocity profile increases
as the value of $n$ increases.  Density profiles 
with larger values of $n$ will approximate a flat rotation curve
over a greater radial extent (in units of $r_{-2}$) 
than profiles with smaller values of $n$ (Figure~\ref{FigVel}c and d). 

\begin{figure}
\includegraphics[scale=0.34,angle=270]{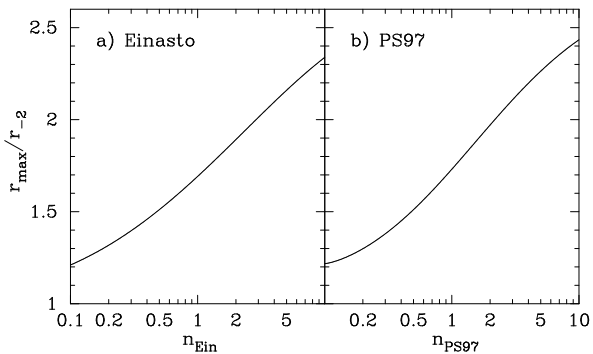}
\caption{
The radius where the circular velocity profile peaks, divided by the 
radius where the logarithmic slope of the density profile equals $-2$,
is shown as a function of the density profile shape $n$ for 
a) the Einasto profile (equation~\ref{SerDen}), and b) Prugniel
\& Simien's profile (equation~\ref{EqPS97}). 
}
\label{FigVcirc}
\end{figure}

The radius $r_{\rm max}$ can be obtained numerically by solving 
the expression for when the derivative of $GM(r)/r$ 
equals zero.  For the Einasto model, this amounts to solving 
\begin{equation}
\gamma(3n,x) = x^{3n} {\rm e}^{-x}/n, 
\end{equation}
with $x = d_n(r_{\rm max}/r_{\rm e})^{1/n}$. 
For the Prugniel \& Simien model, one needs to solve the expression 
\begin{equation}
\gamma(n(3-p),Z) = Z^{n(3-p)} {\rm e}^{-Z}/n, 
\end{equation}
with $Z = b(r_{\rm max}/R_{\rm e})^{1/n}$. 
The results are shown in Figure~\ref{FigVcirc}, with $r_{\rm max}$
normalized against $r_{-2}$.

When $n=3.6$ (and 2.9) in the Prugniel-Simien model --- the average 
profile shape for our galaxy-sized (and cluster-sized) CDM halos, see
Paper~I --- 
$r_{\rm max} \sim$ 2.17 $r_{-2}$ (and 2.10 $r_{-2}$). 
When $n=6$ (and 5.0) in the Einasto model, 
$r_{\rm max} \sim$ 2.21 $r_{-2}$ (and 2.16 $r_{-2}$). 
This can be compared with the ($\gamma =1$) NFW model for
which $r_{\rm max}$ is known to equal $\sim$ 2.16 $r_{-2}$.

\subsection{Concentration and the virial radius, $r_{\rm vir}$}\label{DefinC} 

Given the re-newed application of Einasto's model, which has the
same functional form as S\'ersic's model that is used
by observers in describing projected distributions, it would seem
relevant to inquire if we can also make use of the type of
concentration indices that observers use.  There are two flavors.

The first is a ratio of two radii.  While this may sound qualitatively
similar to the NFW concentration, it is in fact fundamentally
different.  A classical example would be the ratio between the radii
containing 50\% and 25\% of a galaxy's total light (Fraser 1972). 
In the case of a universal density profile, with only a radial scale
and a density scale, the ratio of radii containing 50\% and 25\% of
the total (asymptotic) mass would always be exactly the same.  That
is, if we were to use the new concentration index $r_{\rm e}/r_{-2}$,
then if $n$ is assumed to be constant (i.e., if a universal profile
exists), this ratio will be the same for every profile (see
equations~\ref{SerR2} and \ref{PSR2}). 
A similar example comes from Butcher \& Oemler (1984) who defined a
galaxy cluster concentration index $C = \log(R_{60}/R_{20})$, where
$R_x$ is the radius enclosing $x$\% of the galaxies in a cluster. 

The second type of concentration index is a ratio of flux, compared to
a ratio of radii, within two specified radii (Okamura, Kodaira, \&
Watanabe 1984).  An example is the flux within the radius containing
half an object's total light divided by the flux within one third of
this radius (Trujillo, Graham, \& Caon 2001).  But again, if the
density profiles are universal, then such concentration indices will
always have the same value.  It is because real galaxies do not have
universal light-profiles, i.e.\ a range of S\'ersic ($R^{1/n}$)
indices are observed, that such concentrations indices work.
%
% Intriguingly, such galaxy concentration is known to correlate with a
% galaxy's supermassive black hole mass $M_{\rm bh}$ just as strongly as
% the $M_{\rm bh}$--$\sigma$ relation (Ferrarese \& Merritt 2000;
% Gebhardt et al.\ 2000), and with the same small degree of scatter
% (Graham et al.\ 2001).
%%
% Due to this relation's direct applicability to bulges in disk
% galaxies, in addition to elliptical galaxies, it has recently been used
% to acquire the most accurate to date estimate of the local
% supermassive black hole mass function (Graham et al.\ 2005).

Abraham et al.\ (1994) used a galaxy concentration index defined as
the ratio of flux within the radius containing an object's total light
(rather than half its light) divided by the flux within 1/3 of this
radius.  Now because, in general, galaxies do not have well-defined
edges but rather their light slowly peters out into the noise of the
sky-background flux, deeper and deeper exposures yield increasingly
larger total radii, and a flux ratio that tends to 1 for every galaxy.
But because of the limited aperture sizes Abraham et al.\ (1994) used
to define the total galaxy light, they obtained values different than 1.
The quantity they measured was thus a function of not only the
light-profile shape, but how many $R_{\rm e}$ they sampled in their
largest aperture, as discussed in Graham, Trujillo, \& Caon (2001).

In a somewhat similar manner, the NFW concentration works because it
too is dependent on the background noise, specifically, the mean
matter density of the universe.
The virial radius, $r_{\rm vir}$, is used to quantify the density of
dark matter halos relative to the background.  It is defined as the
radius of a sphere containing an average matter density that is some
specific number greater than the mean matter density in the universe.
In Paper~I we reported that our simulated halo profiles were 
computed using a value of 368.  It is however common to also see a value of 
$\sim$337 when $\Omega_{\rm baryon} = 0.3,
\Omega_{\Lambda} = 0.7$, and $h=0.7$ (e.g., Bryan \& Norman 1998). 
Before the cosmological constant became fashionable, a
value of $\sim$178 was used for the flat Einstein de Sitter universe.

% Note: In Eke, Cole \& Frenk (1996, their Fig.1), one will see a number
% of $\sim$102 at $\Omega=0.3$ for the $\Omega+\Lambda=1$ curve, not
% 337.  This is because it is in units of the larger critical density,
% not the mean matter density.  This number also agrees with the value
% from $\sim 178{\Omega_m=0.3}^{0.45}$ (e.g., Macci\`o et al.\ 2003,
% their equation~1.1).  Dividing 102 by 0.3 gives 340.

\begin{figure*}
\includegraphics[scale=0.97,angle=270]{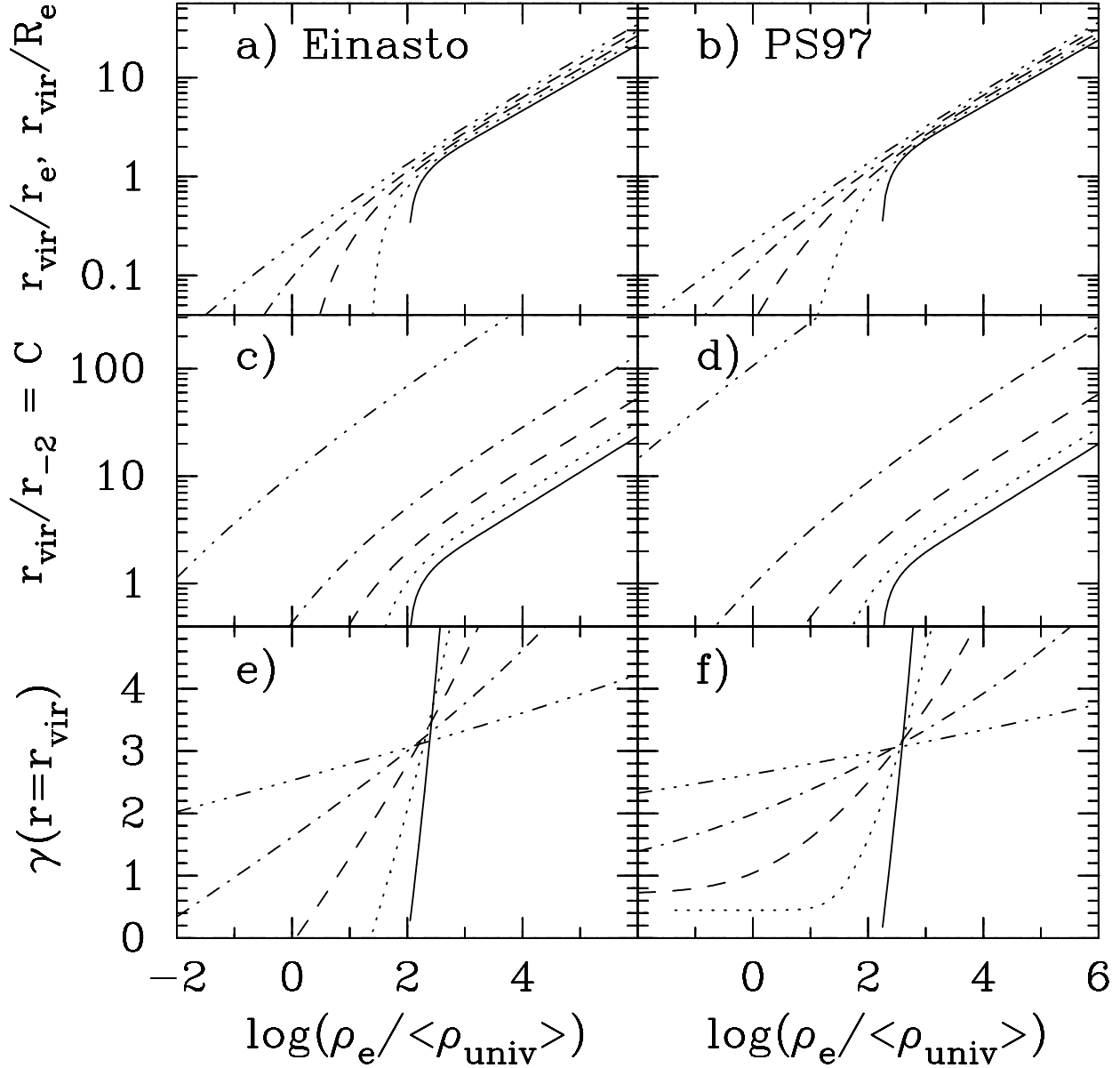}
\caption{
Panels a) and b) show 
the virial radius (normalized against the effective radius 
$r_{\rm e}$ and $R_{\rm e}$, respectively), 
as a function of the scale-density $\rho_{\rm e}$ relative
to the average background density of the universe for Einasto's 
model (equation~\ref{SerDen}) and the Prugniel-Simien model
(equation~\ref{EqPS97}), respectively. 
Panels c) and d) are similar to panels a) and b) but with the 
virial radius normalized by the radius where the logarithmic 
slope of the density profile equals $-2$. 
Panels e) and f) show the associated, negative, logarithmic 
slope of the density profile at the virial radius, denoted by 
$\gamma_{\rm vir}$. 
$n=0.5$ (solid line); $n=1$ (dotted line); $n=2$ (dashed line); 
$n=4$ (dot-dash line); $n=10$ (triple-dot-dash line). 
}
\label{Conc}
\end{figure*}

Using equations~\ref{Sermass} and \ref{PSmass} for the mass profile, 
one can (numerically) solve the following expression to obtain the
virial radius in units of the scale radius $r_{\rm e}$ and $R_{\rm e}$
from the Einasto and Prugniel-Simien model, respectively. 
\begin{equation}
3M(r_{\rm vir})/(4\pi {r_{\rm vir}}^3) = 337\langle \rho_{\rm universe}\rangle. 
\end{equation}
Halos do of course extend beyond $r_{\rm vir}$ (e.g., Macci\'o, Murante, \&
Bonometto 2003; Prada et al.\ 2005).  The results are shown in the top
panels of Figure~\ref{Conc}, where one can see how the virial radius
is monotonically related to the density contrast between $\rho_{\rm
e}$ and $\langle \rho_{\rm universe} \rangle$.
As $r_{\rm vir}$ increases beyond the effective radius, the incomplete
gamma function in the mass profiles starts to asymptote to a constant
value and the slope in the figures tends to 1/3.
%
% Normal NFW concentrations are in the range 5 to 25. 

From the relations connecting $r_{\rm e}$ and $R_{\rm e}$ with
$r_{-2}$ (equations~\ref{SerR2} and \ref{PSR2}), one can obtain the
virial radius in units of $r_{-2}$.  This is shown in the middle
panels of Figure~\ref{Conc}.  The ratio $r_{\rm vir}/r_{-2}$ is
referred to by modelers as the {\it concentration parameter}.  It does
not refer to the curvature or shape of the profiles, as observers
might initially think, but is a measure of the density contrast of the
halo relative to the average background density of the
universe\footnote{Although these diagrams were created using an
over-density factor of 337, this actual choice does not affect the
(modeler's) concentration parameter's ability to act as a surrogate
for the density scale $\rho_{\rm e}$ or $\rho_{-2}$.}.
Obviously, if one did not wish to use the virial radius (see Macci\`o,
Murante, \& Bonometto 2003 for an alternative prescription), then a
similar `concentration parameter' can be defined in terms of
$\rho_{\rm e}/\langle\rho_{\rm universe}\rangle$. 
The slope at $r=r_{\rm vir}$ is shown in the lower panels of
Figure~\ref{Conc} as a function of 
$\rho_{\rm e}/\langle\rho_{\rm universe}\rangle$.

%      note: the curves in panels c) and d) are simply offset
% vertically from those in panels a) and b). 

% xxx What am I missing?  That is, why did people even bother with this
% virial radius to measure the so-called concentration?  rather than
% just use $\rho_{\rm e}/\langle \rho_{\rm univ} \rangle$ or
% $\rho_{-2}/\langle \rho_{\rm univ} \rangle$
%
% @ Maccio03 and Prada05 show that there is no physical justification
% for using this definition. It also causes a lot of confusion since
% rvir and c evolve with (z+1) even if the halo is perfectly
% stationary. And it makes many people think of halos as spheres which
% end at rvir and continuously accrete material (and subhalos, angular
% momentum etc.) at this ``boundary'' but most of what they see are
% effects of increasing the window size with time. It would be great to
% forget about rvir and shift towards using a scale radius and the
% density at this radius as you suggest (both in physical not comoving
% units).

If one thinks of dark matter halos as icebergs, which can be lowered
and raised relative to the background density of the universe, then 
profile universality means that one can use the offset between 
$\langle \rho_{\rm universe} \rangle$ 
and either $\rho_{e}$ or $\rho_{-2}$ or $\rho^{\prime}$ 
as a measure of `concentration'. 
But if a range of profile shapes exists, i.e.\ different $n$ ($\alpha$ 
in the notation of Navarro et al.\ 2004), the difference between 
$\rho_{e}$ and $\rho_{-2}$ and $\rho^{\prime}$ will depend on the profile shape.
What this means is that the concentration one measures will 
depend on where one samples the halo's density. 
This is important because the halo density is thought to reflect the
mean density of the universe when the halo formed, and is thus a
measure of the collapse redshift of the halo.

% xxx We need more halos to do the following: 
% 
It would be of value to explore whether or not the
use of $\rho_{\rm e}$ and $M(r_{\rm e})$ (and $M(R_{\rm e})$ in the case
of the Prugniel-Simien model), rather than $\rho_{-2}$ and $M_{\rm
vir}$ may account for some of the scatter in diagrams plotting
concentration versus halo mass, or equivalently, scale radius versus
scale density. 

Finally, we note that the use of a Petrosian-style radius (Petrosian
1976; Graham et al.\ 2005), such that the mean density inside of some
radius divided by the density at that radius equals some constant
value, is not suitable in the case of structural homology.  This is
because such a radius will be equal to the same fractional number of scale
radii ($r_{\rm e}$ or $r_{-2}$) for every halo.  That is, a
Petrosian-like radius will just be a re-expression of the scale
radius.

\section{On the power-law nature of $\rho/{\sigma}^3$} \label{SecCube}

% Arad et al.\ (2004) have computed the exact local distribution function.

% xxx Henriksen (2006). 

There has been recent interest in the the pseudo phase-space density profiles 
represented by $\rho(r)/{\sigma(r)}^3$. 
This quantity appears to be well approximated by a power-law $r^{-\alpha}$,
with $\alpha \approx 1.94$ (Taylor \& Navarro 2001; Ascasibar et al.\
2004; Rasia, Tormen, \& Moscardini 2004; Sota et al.\ 2006; Barnes et al.\ 2006).  
Independently of any model,
Dehnen \& McLaughlin (2005) found a best-fit value of $\alpha =
1.92\pm0.01$ using the halos A09-F09 and G00-G03, which we studied 
in Paper~I. 

\begin{figure}
\includegraphics[scale=0.62]{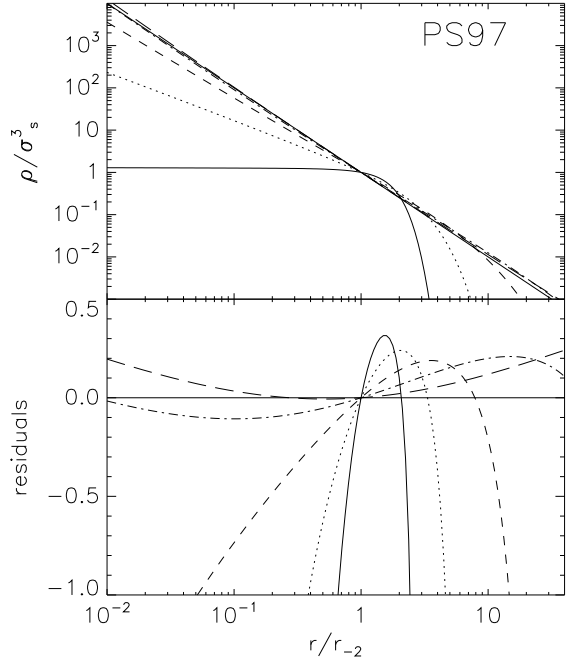}
\vskip30pt
\caption{
Prugniel-Simien density profile $\rho(r)$ (equation~\ref{PSNav})
divided by the cube of its spatial (i.e., not projected) velocity
dispersion profile $\sigma_s(r)$ (Terzi\'c \& Graham 2005, their
equation~28).
The curves are such that: $n=0.5$ (solid line), $n=1$ (dotted), $n=2$
(short-dashed), $n=4$ (dash-dot), $n=10$ (long-dash).
The curves asymptotically approach a line having slope $-2$ (shown by
the solid straight line) as $n \rightarrow \infty$.  
One obtains the same asymptotic behavior
using the projected velocity dispersion profile.
The lower panel shows the difference between the curved profiles and
the line of slope $-2$, divided by the density of the curved profiles.
}
\label{DenSigPS97}
\end{figure}

Figure~\ref{DenSigPS97} shows the ratio $\rho(r)/{\sigma(r)}^3$ for the
Prugniel-Simien density profile (equation~\ref{EqPS97}) coupled with
its spatial (i.e., not projected) velocity dispersion profile given in
Terzi\'c \& Graham (2005, their equation~28).   
Isotropy in velocity space has been assumed. 
As can be seen, the profiles are not exactly featureless
power-laws, but for $n \gtrsim 4$ the departure from a power-law, 
over the radial range shown, is less than about 20\%. 

\begin{figure}
\includegraphics[scale=0.62]{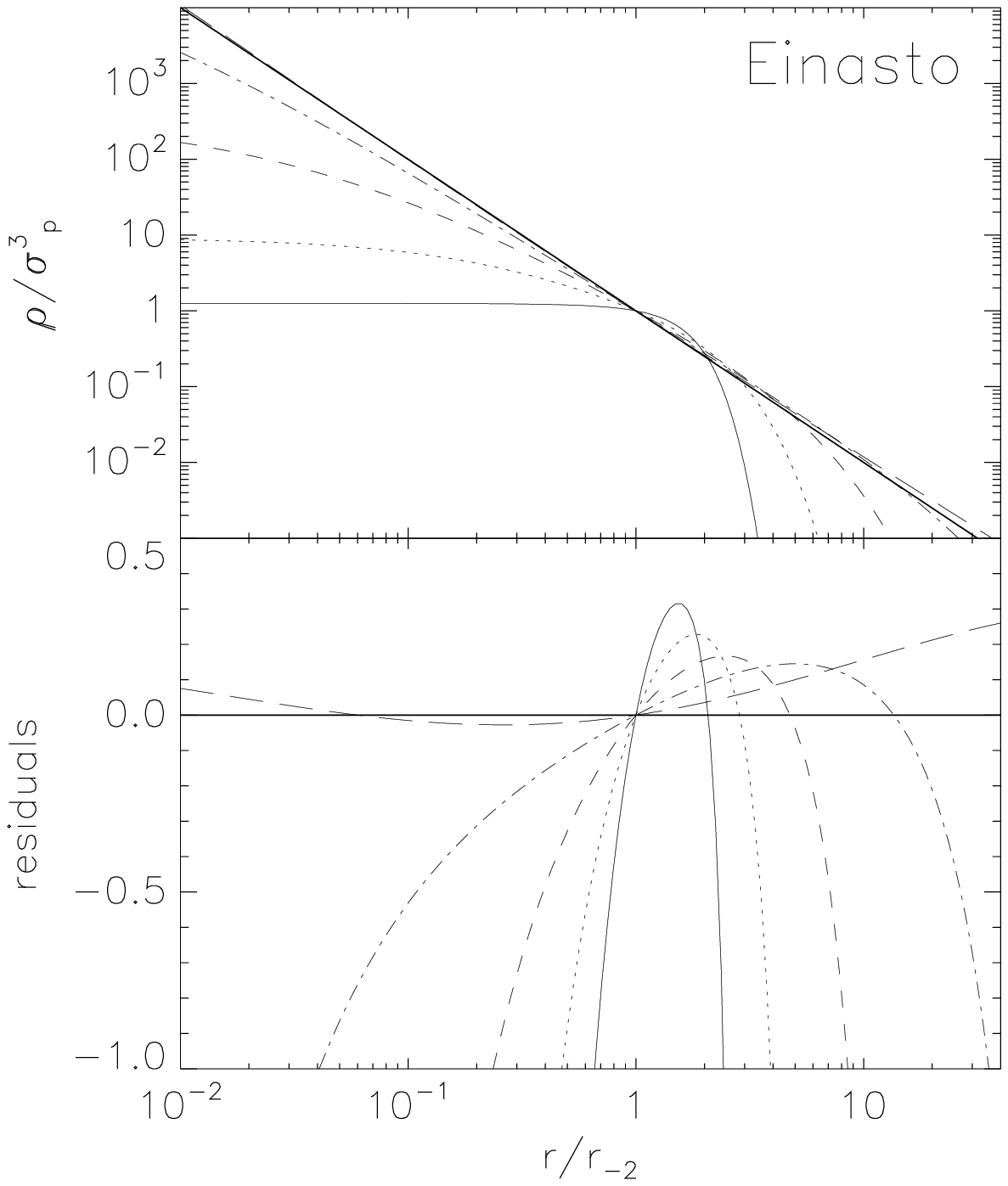}
\vskip30pt
\caption{
Same as figure~\ref{DenSigPS97} except that Einasto's 
density profile (equation~\ref{SerNav}) has been used here. 
}
\label{DenSigSer}
\end{figure}

For the Einasto density profile (equation~\ref{SerNav}),
the spatial velocity dispersion profile can be obtained by 
integrating the isotropic Jeans equation of hydrostatic equilibrium 
\begin{equation} 
{\sigma_{\rm s}}^2 (r) = {G \over {\rho(r)}} \int_r^{\infty} \rho({\bar r})
{{M({\bar r})} \over {\bar r}^2} {\rm d}{\bar r}. 
\end{equation}
Expressing $r_{\rm e}$ in terms of $r_{-2}$ (equation~\ref{SerR2}) in
the mass profile $M(r)$ (equation~\ref{Sermass}), one obtains
\begin{eqnarray} 
{\sigma_{\rm s}}^2 (x) & = & 
\frac{GM_{\rm tot}}{r_{-2}} \frac{(2n)^{1+n}}{2} \frac{{\rm e}^x}{\Gamma(3n)} \nonumber \\
& & \hskip-20pt \times 
\int_x^{\infty} {\bar x}^{-n-1} {\rm e}^{-{\bar x}} \gamma \left(3n,
{\bar x} \right) {\rm d}{\bar x},
\end{eqnarray}
where ${\bar x} = d_n({\bar r}/r_{\rm e})^{1/n} = 2n({\bar r}/r_{-2})^{1/n}$. 
Integration to infinity for this 
expression\footnote{Equation~23 in Cardone et al.\ (2005) for the
velocity dispersion profile contains a typo such that the term in
the exponential should be $(+2/\gamma)(r/r_{-2})^{\gamma}$ rather than
$(-2/\gamma)$, where their $\gamma$ equals $1/n$. Their figure~5 is
however correct.}
is avoided by making the substitution ${\bar
x}=(x/\cos \theta)$, such that ${\rm d}{\bar x}/{\rm d}\theta =
x\sin\theta/\cos^2\theta$, giving
\begin{eqnarray}
{\sigma_{\rm s}}^2 (x) & = &
\frac{GM_{\rm tot}}{r_{-2}} \frac{(2n)^{1+n}}{2} \frac{{\rm e}^x x^{-n}}{\Gamma(3n)} \nonumber \\
 & & \hskip-70pt \times 
\int_0^{\pi/2} \tan\theta \cos^n \theta \hskip3pt {\rm e}^{-x/\cos\theta}
\hskip2pt \gamma \left(3n, \frac{x}{\cos\theta} \right) {\rm d}\theta.
\end{eqnarray}
Figure~\ref{DenSigSer} shows $\rho/{\sigma_s}^3$ for the Einasto 
density profile. 

\begin{figure}
\includegraphics[scale=0.62]{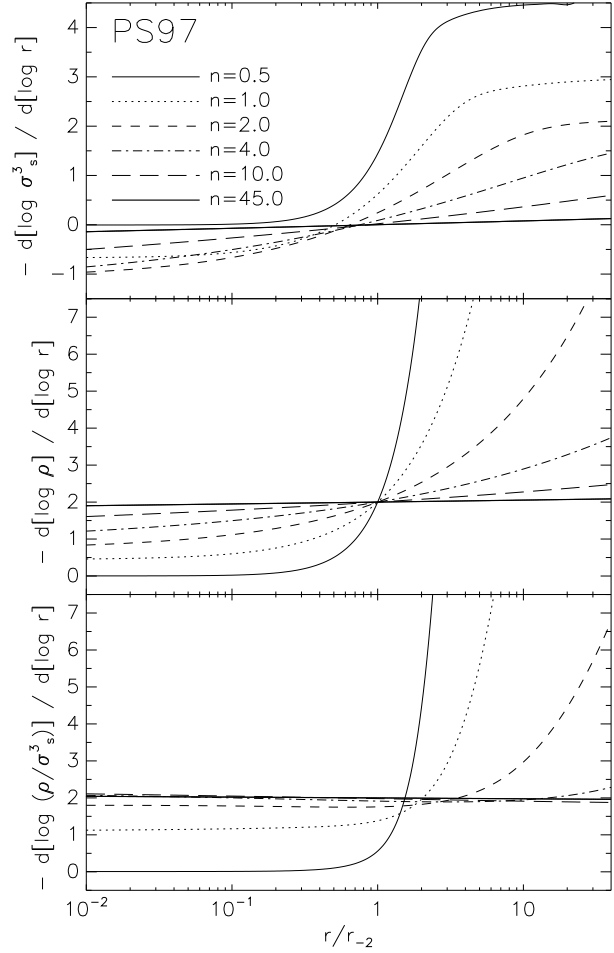}
\vskip30pt
\caption{
Negative, logarithmic slopes associated with the Prugniel-Simien model
(equation~\ref{PSNav}).  
The curves in the lower panel equal  
the curves in the middle panel minus those in the upper panel. 
See section~\ref{SecCube} for details. 
}
\label{PStri}
\end{figure}

\begin{figure}
\includegraphics[scale=0.62]{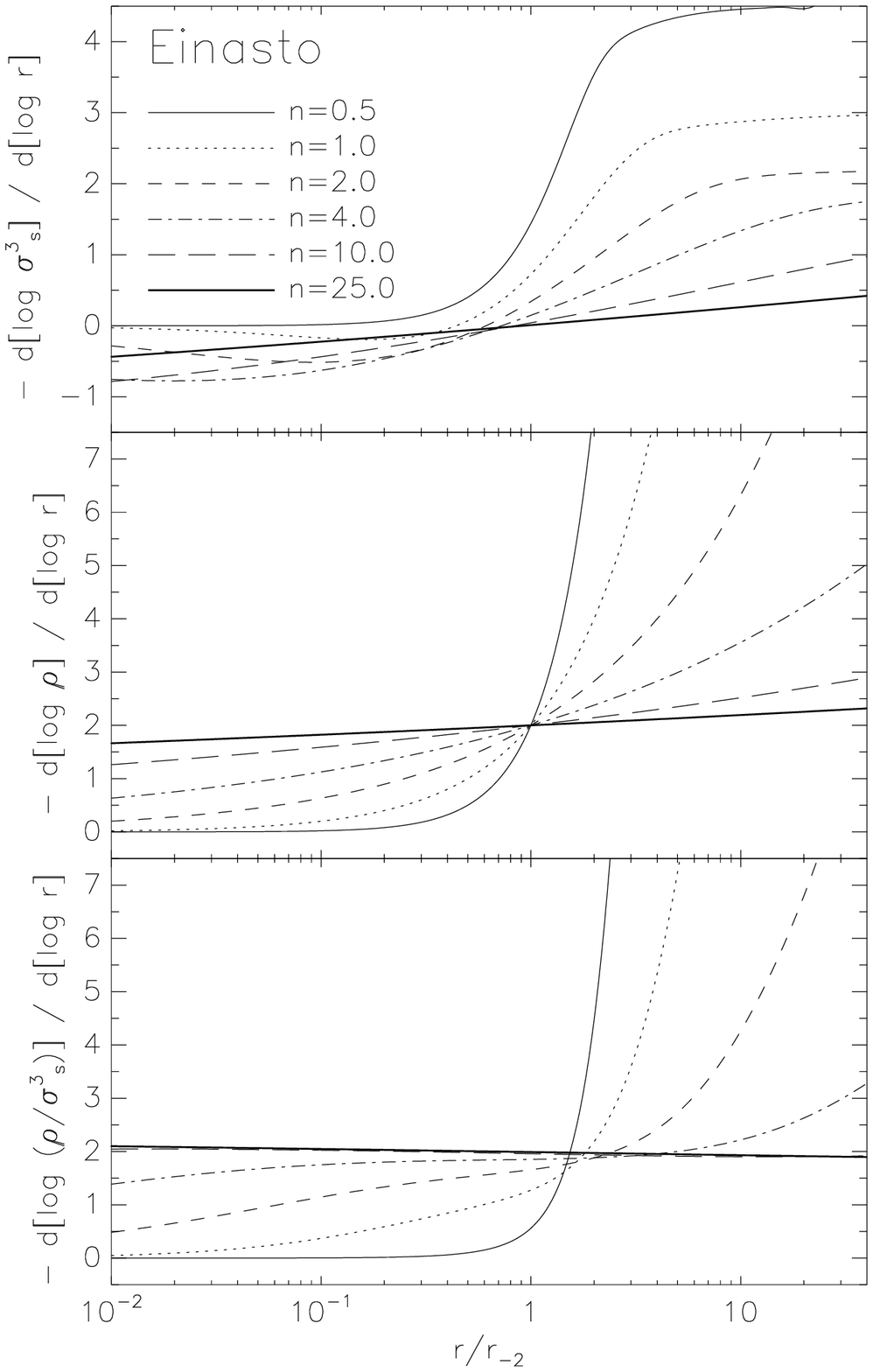}
\vskip30pt
\caption{
Same as figure~\ref{PStri} except that Einasto's 
density profile (equation~\ref{SerNav}) has been used here. 
}
\label{Sertri}
\end{figure}

From the residual profiles in Figures~\ref{DenSigPS97} and
\ref{DenSigSer}, one can see that, over the radial range $0.1 <
r/r_{-2} < 10$, a slightly shallower slope than $-2$ exists for $4<n<10$
(Einasto case, Figure~\ref{DenSigSer}) and $2<n<10$ (Prugniel-Simien
case, Figure~\ref{DenSigPS97})\footnote{Had 
the slope been exactly $-2$ in Figures~\ref{DenSigPS97} and \ref{DenSigSer}, 
it would have implied an outer logarithmic slope in the density 
profile that decayed in an oscillary manner about a value of $-2$ 
(Dehnen \& McLaughlin 2005, their figure~2), at odds with the data
in Figure~3 from Paper~I}. 
Although at $r=r_{-2}, -{\rm d}\log \rho/{\rm d}\log r \equiv \gamma = 2$, 
for $n_{\rm Ein}=6, \gamma_{\rm Ein}(0.1r_{-2})=1.36$ and $\gamma_{\rm Ein}(10r_{-2})=2.94$, and
for $n_{\rm PS}=3, \gamma_{\rm PS}(0.1r_{-2})=1.62$ and $\gamma_{\rm PS}(10r_{-2})=2.56$.
The slope of the density profile {\it does} therefore 
change over this radial range.
This can be appreciated in Figures~\ref{PStri} and \ref{Sertri} 
which show the negative, logarithmic slope of the (cube of the) 
velocity dispersion profile, the density profile, 
% (a radially zoomed in version of Figure~\ref{FigGam}), 
and $\rho/{\sigma_s}^3$. 
For large values of $n$, $\rho/{\sigma_s}^3 \approx r^{-2}$. 

% At $r_{-2}$, the slope of the density profile is $-2$, not $-3$ !!xxx
%
% As $n \rightarrow \infty$, the profiles in
% Figures~\ref{DenSigPS97} and \ref{DenSigSer} appear to approach
% a power-law with slope $-2$, i.e., $\rho/{\sigma_s}^3 \rightarrow r^{-2}$.
% 
% This arises from the density profiles tending to $\rho(r) \propto
% r^{-3}$ (see equations~\ref{SerGam} and \ref{PSGam} as $n \rightarrow
% \infty$), and the velocity dispersion profiles tending to power-laws
% such that $\sigma_s(r) \propto r^{-1/3}$.
% % [XXX check this is true, which would confirm the above point xxx].
% % 
% This, however, does not explain the near power-law nature of
% $\sigma/\rho^3$, near $r=r_{-2}$, for our halos.  The density profiles
% under investigation have optimal shape parameters $n$ around 6
% (Einasto model) and 3 (Prugniel-Simien model).  They are therefore
% not power-laws with slopes equal to $-2$, but possess significant
% curvature, as is obvious from Figure~3 of Paper~I..

\section{Density profile slopes} \label{SecSlope}

% empirical models that had steep asymptotic inner density
% profiles --- with slopes between 1 and 1.5 (e.g., Navarro et al.\
% 1997; Moore et al.\ 1998; Jing \& Suto 2002; Power et al.\ 2003; Rasia
% et al.\ 2004) --- and the ``core crisis'' has persisted.
% 
% From the Jeans equation, 
% M\"ucket \& Hoeft (2003) argue that logarithmic slopes shallower than $-1$ 
% are expected as $r \rightarrow 0$ for spherically symmetric systems. 
% 
% Stoehr et al.\ (2002) introduced a new model that fitted simulated
% $\Lambda$CDM data equally as well as the (1, 3, $\gamma$) model (Stoehr
% 2006; Diemand et al.\ 2005) but had an inner slope of zero that
% appeared to resolve the ``core crisis''.  Navarro et al.\ 2004 seemed
% to dismiss the model for it's strange behavior ??

De Blok (2005) has argued that the inner profile slopes of simulated,
dark matter halos are inconsistent with observations of dark matter
dominated galaxies (Moore et al.\ 1999; Salucci \& Burkert 2000;
Marchesini et al.\ 2002; de Blok, Bosma, \& McGaugh 2003; Gentile et
al.\ 2006; Goerdt et al.\ 2006). 
He reports that the inner density profiles of low surface
brightness (LSB) galaxies have logarithmic slopes significantly
shallower than $-1$ at a radius of 0.4 kpc.  This is important because
it suggests a possible problem with hierarchical $\Lambda$CDM
simulations of dark matter halos, which, at least from (1, 3,
$\gamma$) model fits, typically have inner slopes steeper than $-1$.

While there is presently no consensus as to why such a disagreement
exists, some of the apparent discrepancy may arise from either 
baryonic processes which modify the dark matter profile (e.g., 
Mashchenko, Couchman, \& Wadsley 2006), or from systematic 
biases in measuring inner slopes from HI and H$\alpha$ long-slit
observations (van den Bosch et al.\ 2000; Swaters et al.\ 2003a;
Spekkens, Giovanelli, \& Haynes 2005, but see de Blok 2003).  For
example, non-circular motions can make galaxies appear less cuspy than
they really are.  Significant, in the sense of non-zero, non-circular
motions are indeed present in many LSB galaxies where high resolution
2D velocity fields are available (e.g.,
% Schoenmakers, Franx, \& de Zeeuw 1997: non-LSB NGC 2403 and 3198
Swaters et al.\ 2003b; Blais-Ouellette et al.\ 2004; Coccato et al.\
2004; Simon et al.\ 2005).  However, on their own, these do not
explain the observed difference in slope (de Blok et al.\ 2003;
Gentile et al.\ 2006).  But in combination with gas pressure support
and projection effects, Valenzuela et al.\ (2005) argue that this may
account for the relatively shallow slopes in observations.  Using
extensive simulations of observing and data processing techniques,
Spekkens et al.\ (2005) also report how measurements from long-slit,
optical spectra of halos with inner slopes of $-1$ can result in
``observed'' slopes consistent with values around $-0.25 \pm0.15$.  The
apparent success of the flat-core Burkert (1995) model may then be an
artifact of observational biases.
Higher-resolution gamma-ray studies of dark-matter dominated galaxies 
may, in the future, help to resolve the current cusp-core
controversy (e.g., Profumo \& Kamionkowski 2006; Lavalle et al.\ 2006). 

The inward extrapolation of simulated density profiles using empirical
models which have steep (asymptotic) inner power-laws may also be
partly responsible for the mismatch (see., e.g., Kravtsov et al.\ 1998). 
As noted by Navarro et al.\ (2004) and Stoehr (2006), empirical models 
with shallow inner slopes, such as Einasto's model, not only match 
the simulated data down to 0.01 $r_{\rm vir}$ but could potentially
resolve the apparent dilemma at smaller radii (but see Diemand et al.\
2005, who find a slope of $-1.25$ at 0.001 $r_{\rm vir}$ in a
highly-resolved, cluster-sized halo).  That is, $\Lambda$CDM cosmology
and the various $N$-body simulations themselves may in fact be fine,
but the empirical models used to parameterize the CDM halos may fail
at small radii.  
% On the other hand, the simulations may require modifications 
% that could shed new insight into the nature of our universe.
%% Xiao et al.\ (2005)  astro-ph/0508436

Here we examine the slope of the various empirical models,
and compare these with observations of real, dark matter dominated 
galaxies.

%   rho ~ 1/x^gamma / (1+x)^(3-gamma),  where x = r/r_s, then
%    d rho/dx = - gamma rho/x^gamma - (3-gamma) rho/(x+1)
%  hence
%     gamma_(1,3,gamma)(r) = - d ln rho / d ln x
%                          = gamma + (3-gamma) x/(x+1)
%                          = gamma + (3-gamma) r/(r+r_s)

\begin{figure}
\includegraphics[scale=0.55,angle=270]{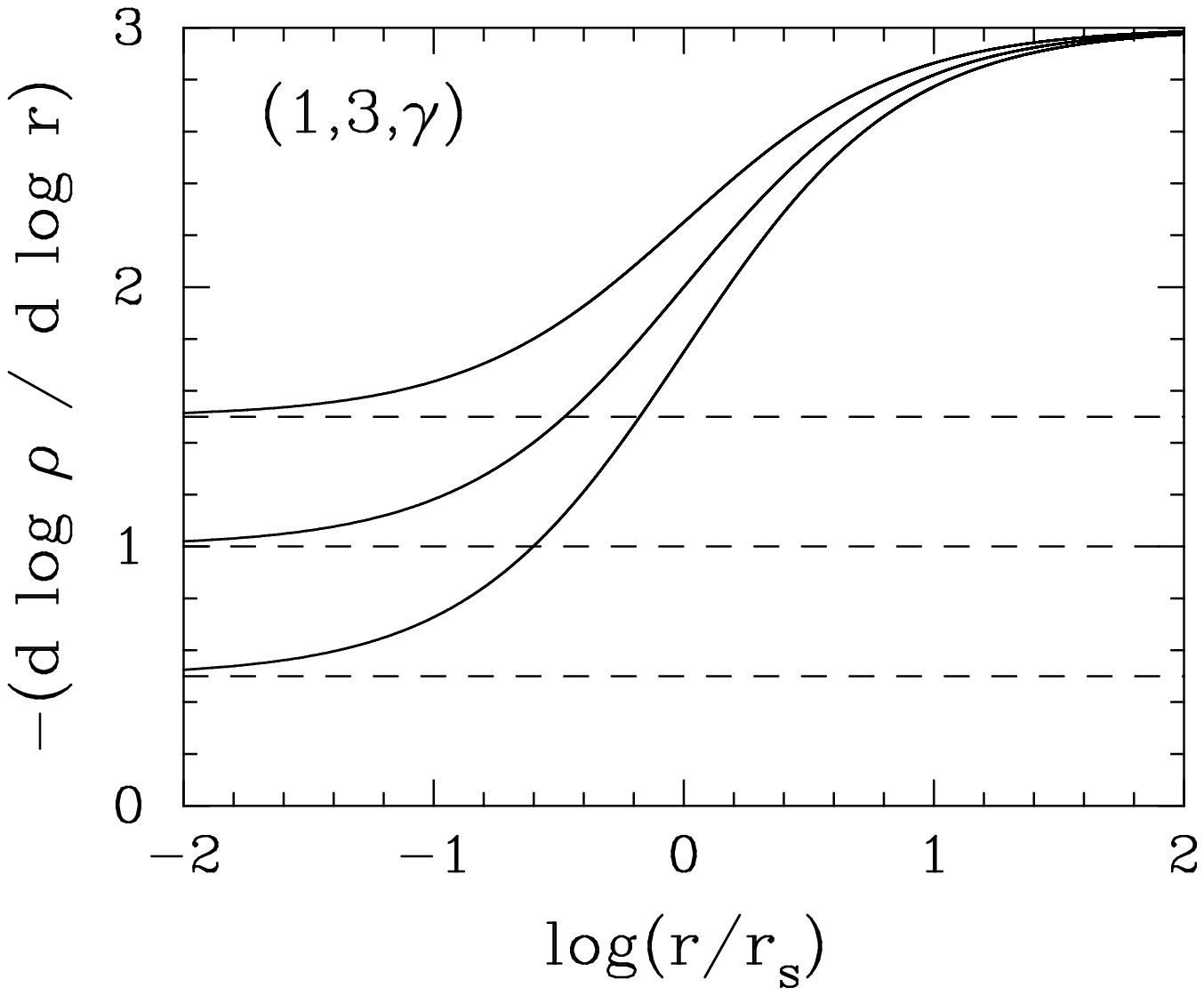}
\caption{
Negative logarithmic slope of the (1, 3, $\gamma$) model (equation~\ref{EqNFW}) 
when $\gamma$=0.5, 1.0, and 1.5 (bottom, middle, and top curves, respectively).
}
\label{NFWslop}
\end{figure}

In the case of the (1, 3, $\gamma$) model (equation~\ref{EqNFW}), the
slope is given by
\begin{equation}
\gamma_{(1,3,\gamma)}(r) \equiv \frac{-{\rm d}[\log \rho(r)]}{{\rm d}\log r}
 = \gamma + (3-\gamma)/(1+r_s/r). 
\end{equation}
For small values of $r/r_s$, 
\begin{equation}
\gamma_{(1,3,\gamma)}(r) \approx \gamma + (3-\gamma)\frac{r}{r_s}
 \hskip30pt \mbox{$r << r_s$}, 
\end{equation}
which, as expected, asymptotically approaches $\gamma$ as $r \rightarrow 0$. 
Figure~\ref{NFWslop} shows the negative, logarithmic slope as a
function of radius for a sample of (1, 3, $\gamma$) models with
$\gamma=$0.5, 1.0, and 1.5.  One can see that the negative logarithmic
slope of the profiles are practically equal to $\gamma$ at $r \lesssim
0.01 r_s$.  What should also be realized is that, although the
(1, 3, $\gamma$) models do have continuously curving slopes from
0.01 to 1 $r_{\rm vir}$, they don't
have the correct continuously curving slope to match the CDM halos as well
as Einasto's model can or the Prugniel-Simien model can. 

\begin{figure}
\includegraphics[scale=0.47,angle=270]{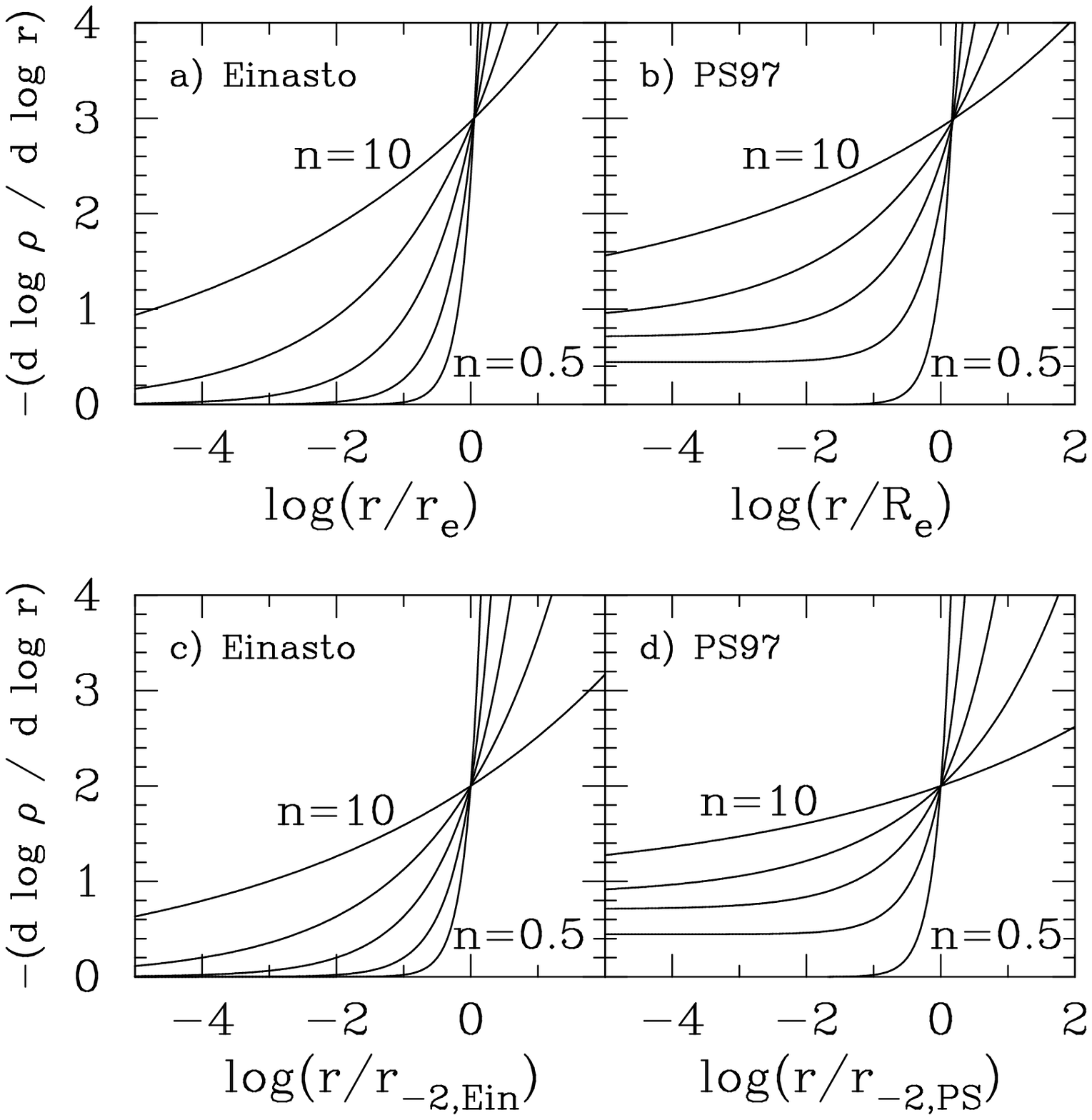}
\caption{
Panel a) Negative logarithmic slope of the Einasto $r^{1/n}$ profile
(equation~\ref{SerGam}) as a function of normalized radius $r/r_{\rm
e}$ for different values of $n=0.5$, 1, 2, 4, and 10.  Panel b)
Negative logarithmic slope of Prugniel \& Simien's density profile
(equation~\ref{PSGam}) as a function of normalized radius $r/R_{\rm
e}$.
Panels c) and d) are the same as a) and b) except that the radius has
now been normalized at $r_{-2,{\rm Ein}}$ and $r_{-2,{\rm PS}}$,
respectively (see equations~\ref{SerR2} and \ref{PSR2}).
As $r \rightarrow 0$, $\gamma_{\rm Ein} \rightarrow 0$ while
$\gamma_{\rm PS} \rightarrow p$. 
}
\label{FigGam}
\end{figure}

The negative logarithmic slope of Einasto's model
(equation~\ref{SerDen}) is given by
\begin{equation}
\gamma_{\rm Ein}(r) 
 = -\left( \frac{r}{\log {\rm e}}\right) 
           \frac{{\rm d}[\log \rho(r)]}{{\rm d}r} 
 = \frac{d_n}{n}\left(\frac{r}{r_{\rm e}}\right)^{1/n}, 
\label{SerGam}
\end{equation}
which is approximately $3(r/r_{\rm e})^{1/n}$ for $n\gtrsim 1$ (see
Figure~11 in Paper~I). 
%  ; c.f.\ equation~4 in Graham et al.\ 2003).  
One can also see that when $r = r_{\rm e}$, the
negative logarithmic slope of the density profile is approximately 3.
From Figure~3 of Paper~I it is clear that $r_{\rm e}$ will occur at a
large radius.  For fixed values of $n$, Figure~\ref{FigGam}a shows how
the negative logarithmic slope decreases monotonically as the radius
$r$ decreases.
%
% In passing we note that the model of Burkert (1995, his equation~2) 
% also asymptotes to a central logarithmic slope of zero. 

The negative logarithmic slope of Prugniel \& Simien's model
(equation~\ref{EqPS97}) is given by 
\begin{equation}
\gamma_{\small {\rm PS}}(r) 
 = \frac{b_n}{n}\left( \frac{r}{R_{\rm e}} \right)^{1/n} + p.
\label{PSGam}
\end{equation}
From $b_n \approx 2n-1/3+0.009876/n$ for $n\gtrsim 0.5$ (Prugniel \&
Simien 1997), $\gamma_{\rm PS}(r) \approx 2(r/R_{\rm e})^{1/n} +p$,
and thus $\sim (2+p)$ at the effective radius $R_{\rm e}$.  For large
$n$, $p \rightarrow 1$ and $\gamma_{\rm PS}(r_{\rm e}) \rightarrow 3$,
as is the case with $\gamma_{\rm Ein}(r_{\rm e})$.  These profile slopes
are shown in Figure~\ref{FigGam}b as a function of radius for
different values of the profile shape $n$.

Using equations~\ref{SerR2} and \ref{PSR2}, 
one can reformulate the above equations to obtain
% \begin{equation}
% \gamma_{\small {\rm Ein}}(r)
%  = 2\left( \frac{r}{r_{-2}} \right)^{1/n}, 
% \end{equation}
% and
\begin{equation} \label{SerGam2}
\gamma_{\rm Ein}(r) = 2( r/r_{-2,{\rm Ein}} )^{1/n}, 
\end{equation}
and 
\begin{equation} \label{PSGam2}
\gamma_{\small {\rm PS}}(r)
 = (2-p)\left( \frac{r}{r_{-2,{\rm PS}}} \right)^{1/n} + p. 
\end{equation}

As $r \rightarrow 0$, $\gamma_{\rm Ein} \rightarrow 0$, apparently in
fair agreement with the observations of real galaxies reported in, for
example, Simon et al.\ (2003) and de Blok et al.\ (2003), who find a
negative logarithmic slope of $0.2\pm0.2$ (but see
Section~\ref{SecLSB}).  In the case of the Prugniel-Simien model, as
$r \rightarrow 0$, $\gamma_{\rm PS} \rightarrow p$ ($= 1.0 - 0.6097/n
+ 0.05463/n^2$).  Results from Paper~I gave galaxy (and
cluster) profile shapes ranging from $n_{\rm PS}=$ 3.14 to 4.55 
(and from $n_{\rm PS}=$ 
2.19 to 3.47), suggesting a range of central ($r=0$), negative logarithmic 
profile slopes for the Prugniel-Simien model of 0.81--0.87 (and
0.73--0.83).  These slopes are considerably shallower than the mean
($\pm$ standard deviation) value $\gamma = 1.32\pm 0.19$ (and $1.15\pm
0.16$) obtained from the (1, 3, $\gamma$) model fits
(Paper I).  They are also in excellent agreement with
theoretical expectations based on phase-space arguments which suggest
that CDM density profiles should have central cusp slopes equal to
0.75 (Taylor \& Navarro 2001, see also Hansen \& Stadel 2005 and 
An \& Evans 2006).

\clearpage

\subsection{Slope comparison with real galaxies} \label{SecLSB}

For a more meaningful comparison between model halos and observations of
real galaxies, observers and modelers should report profile slopes as
a function of radius 
% (e.g., Diemand et al.\ 2004, their table~4 and figure~6) 
and perform their comparisons at the same radii.  Remarks in the
literature that higher resolution 
% (and thus it is inferred better and more relevant) 
rotation curves tend to show the greatest departure from an inner
logarithmic slope of $-1$ (or $-1.5$) are somewhat beguiling.  Because
such measurements of the inner profile slope in real galaxies were
often made at radii smaller than those typically probed by
$\Lambda$CDM simulations, they do not provide a particularly strong
constraint or check on the simulations.  They do however provide a
check on any empirical fitting functions whose inward extrapolation
does not fall below some fixed slope, such as $-1$.
Addressing this issue, de Blok et al.\ (2005) has compared real and
simulated systems at 0.4 kpc. He found that the density profiles
implied by the best-fitting, 3-parameter function used by Hayashi et
al.\ (2004; equation~8 from Rix et al.\ 1997) to model the velocity
profiles of LSB galaxies have slopes which are
inconsistent with a value as steep as $-1$, and thus also with the
average value of $\sim -1.2$ that is typically reported for simulated halos.

\begin{figure}
\includegraphics[scale=0.34,angle=270]{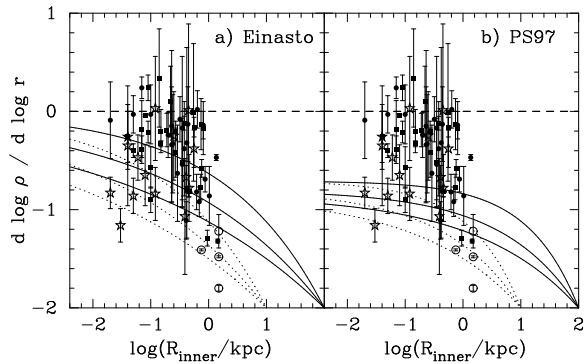}
\caption{
Adaptation (see footnote~\ref{LSBfoot}) of figure~3 from de Blok (2003).
Data points show the logarithmic slope of the density profile
(assuming a minimum stellar disk) at the innermost resolved radius for
a sample of 70 real galaxies.
Open circles: Verheijen (1997); 
filled circles: de Blok et al.\ (2001); 
filled squares: de Blok \& Bosma (2002); 
stars: Swaters et al.\ (2003a). 
Over-plotted in the left panel are the profile slopes from Einasto's 
$r^{1/n}$ model 
(equation~\ref{SerNav}) with $r_{-2}$=10 kpc (dotted lines) and 100 kpc
(solid lines) for $n$=4 (upper curve), 6
(middle curve), and 8 (lower curve).  The right panel shows the same
thing but for the Prugniel-Simien model (equation~\ref{PSNav}) with
$n$=2 (upper curve), 3 (middle curve), and 4 (lower curve).
}
\label{FigLSB}
\end{figure}

For a mean value of $n_{\rm Ein} \sim 6$ (from the Einasto $r^{1/n}$ fits in
Paper~I), a negative logarithmic slope of 0.5 occurs at
$2.4\times 10^{-5}r_{\rm e}$ ($2.4\times 10^{-4}r_{-2}$).  This is
about 10 pc for a galaxy halo with $R_{\rm e}=400$ kpc, and
corresponds to 0.12 arcsec at the distance of the Virgo cluster
(17 Mpc).  
At 0.1 kpc, a typical value at which observers measure the slope of
the mass-density profile in real galaxies (see Figure~\ref{FigLSB}), one
would expect to find a negative logarithmic slope equal to 0.73 for
this halo;
% smaller than expected from the NFW model, 
% but greater than reported for many galaxies. 
in perfect agreement with the mean slope obtained by Simon et al.\
(2005) for a sample of real galaxies.  
At 0.4 kpc, one has $\gamma_{\rm Ein} \sim 0.92$.
If $n=5$ and $R_{\rm e}=200$ kpc, then at 0.4 kpc one has 
$\gamma_{\rm Ein} = 0.85$, and at 0.1 kpc $\gamma_{\rm Ein} = 0.64$, 
consistent with the data from Swaters et al.\ (2003a). 

Figure~\ref{FigLSB} shows the innermost, resolved, logarithmic slope
from the density profiles (assuming a minimum stellar disk) of 70
faint, LSB galaxies thought to be dark matter
dominated (de Blok \& McGaugh 1997; but see the
warning\footnote{Although many are, not all LSB 
galaxies are particularly dark-matter dominated.  For example,
UGC~3137 and UGC~5750 from de Blok \& Bosma (2002) have faint, central
$B$-band surface brightnesses of 24.1 and 23.5 mag arcsec$^{-2}$
respectively, yet their total mass (stars, gas, dark matter) within 4
scale-lengths (=${v_{\rm rot}}^2 4h/G$) divided by their flux within
this radius (equal to 91\% of the total, exponential, disk flux) gives
a solar $M_{\rm tot}/L_R$ ratio of only 13 and 11, respectively.
Typical $M_{\rm tot}/L_B$ ratios for Sa-Sd galaxies, 
within $\sim$4 scale-lengths,  
are 3 to 7 (Roberts \& Haynes 1994).  
% page 119 of Roberts \& Haynes gives the definition of $L_B$.
% page 120 of Roberts \& Haynes gives the definition of $M_{\rm tot}$.
Baryonic processes (Weinberg \& Katz 2002)
might therefore be important here, especially if fractionally more HI 
gas exists in LSB galaxies.}  in Graham 2002), plotted against the
physical radius at which the slope was measured, $R_{\rm inner}$.
This figure has been adapted from de Blok (2003, his
figure~3)\footnote{Figure~\ref{FigLSB} differs slightly from figure~3
in de Blok (2003) because we have included all 15 data points from
Swaters et al. (2003a),
% while de Blok plotted 13 points,
and we have correctly reversed the symbols used to differentiate the
data from de Blok et al.\ (2001) with that from de Blok \& Bosma
(2002).\label{LSBfoot}}.
% 
% Because these galaxies do not all have the same radial scale, $r_{-2}$
% or $r_{\rm e}$, this figure contains a considerable amount of scatter.
% 
In order to compare how well the new density models perform, it is 
necessary to plot several profiles with differing scale
radii --- which amounts to a horizontal shift of the curves in
Figure~\ref{FigLSB}.

Einasto's $r^{1/n}$ model appears capable of matching the data
reasonably well, depending on the combination of scale radius and
profile shape $n$.  However, for the halos studied in Paper~I, bounded by
the curves shown in Figure~\ref{FigLSB}, the best-fitting Einasto
models do not have negative, logarithmic slopes shallower
than $\sim$0.4 at radii $\gtrsim$0.1 kpc.  This is at odds with
roughly half of the galaxies from de Blok et al.\ (2001) and de Blok
\& Bosma (2002), but largely in agreement with the data from Swaters
et al.\ (2003a) and Verheijen (1997).  Clearly, the apparent inconsistency
between the inner profile slope of dark matter halos generated from
$\Lambda$CDM $N$-body simulations and observations of real galaxies is
reduced upon replacement of the NFW model with the (better fitting) Einasto model.
% 
%, and illustrates the dangers of parametric models.  
%
% Past criticisms of $N$-body computer simulations should perhaps have
% been more directed at the NFW-like empirical fitting function.
% 
What is also apparent is that one can expect a range of different
slopes inside of 1 kpc; although this in itself does {\it not} imply a
non-universal density profile.
The largest study to date of (165) low-mass galaxies found inner
logarithmic slopes ranging from $-0.22\pm0.08$ to $-0.28\pm0.06$ for
various subsamples of the data (Spekkens et al.\ 2005).  However,
after extensive testing, these authors concluded that, due to biases
in the analysis of long-slit spectra, the data is in fact consistent
with inner logarithmic slopes ranging from 0 to $-1$.

The extrapolation of the Prugniel-Simien model inside of
$\sim$1 kpc --- the extent to which our simulations (Paper~I) 
provide meaningful 
data --- does not do so well at matching the observations of 
real galaxies (Figure~\ref{FigLSB}b).  This remark does however
overlook the previously noted analysis of Spekkens et al.\ (2005). 
In passing we note that Demarco et al.'s (2003) analaysis of 24 galaxy clusters
using the Prugniel-Simien model yielded a mean inner slope of
$-0.92$. 

Although not explored in this work, the power-law exponent $p$ in
the Prugniel-Simien model can be modified. As we saw in Figure~\ref{FigGam}, as
$r\rightarrow 0$, the negative logarithmic slope of this model tends
to the value $p$.  If one was to reduce the value of $p$, then one
would acquire shallower inner slopes.  As we noted before, 
if one reduced the value of $p$
to zero, then one would obtain Einasto's model.

\begin{figure}
\includegraphics[scale=0.4,angle=270]{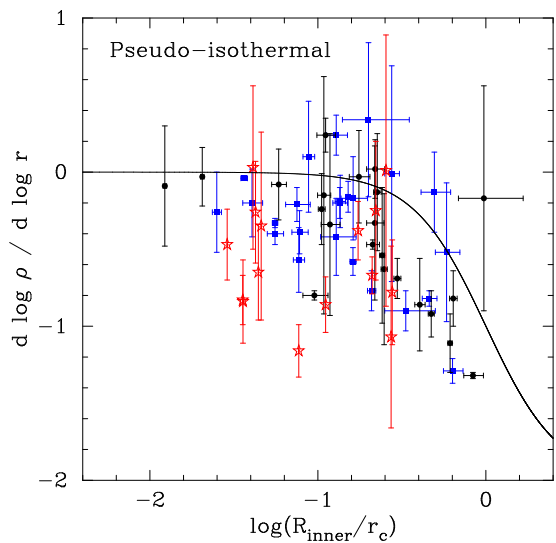}
\caption{
Solid line: logarithmic slope of the pseudo-isothermal model
(equation~\ref{EqIso}). The data points are from observations of real
galaxies (assuming a minimum stellar disk).
Open circles: Verheijen (1997); 
filled circles: de Blok et al.\ (2001); 
filled squares: de Blok \& Bosma (2002); 
stars: Swaters et al.\ (2003a). 
If these galaxies were described by the pseudo-isothermal model,
they would follow the curve. 
}
\label{FigIso}
\end{figure}

Ideally, rather than simply plotting the inner profile slope versus
the radius in kpc at which the slope has been measured, one should
factor in that galaxies possess a range of sizes, i.e.\ scale radii.
For example, 0.4 kpc may correspond to 0.5 scale radii or 0.05 scale radii. 
Although neither the Einasto nor Prugniel-Simien models have yet been
applied to the rotation curve data of the 70 galaxies shown in
Figure~\ref{FigLSB}, a pseudo-isothermal model has been fit to most of
these galaxies.  This simple model can be written as
\begin{equation} \label{EqIso}
\rho(r) = \frac{\rho_0}{1+(r/r_c)^2}, 
\end{equation}
where $r_c$ is the core radius and $\rho_0$ the central density.
Figure~\ref{FigIso} shows the logarithmic slope of this model,
together with data from de Blok et al.\ (2001), de Blok \& Bosma
(2002), and Swaters et al.\ (2003a) for which values of the 
scale radius $r_c$ were 
available.  The scattering of points, rather than following the curve,
suggest that the data do not behave according to the pseudo-isothermal
model, and/or an underestimation of $R_{\rm inner}$ relevant to where
the slope was measured.
Note though that the pseudo-isothermal model is an extreme model with
a somewhat large, flat inner density profile: models based on recent
observations favor a slightly steeper slope of $-0.2\pm$0.2 (e.g.\ de
Blok et al. 2001, 2003), while others find a slope scattered around
$-0.73\pm0.44$ (e.g.\ Simon et al.\ 2005; Swaters et al.\ 2003a).
Steeper cusp models would provide a better fit to the data shown in
Figure~\ref{FigIso}.
It would be of interest to obtain the best-fitting Einasto radii
$r_{\rm e}$ and profile shapes $n_{\rm Ein}$ for these galaxies, which would
allow one to compare how well the observed inner slopes correlate with
$R_{\rm inner}/r_{\rm e}$.

\section{Summary} \label{SecSum}

We have provided expressions to relate the half-mass radii of the
Einasto and Prugniel-Simien models to a) the radius, $r_{-2}$, where
the logarithmic slope of the density profile equals $-2$, b) the virial
radius, $r_{\rm vir}$, and c) the radius where the associated circular
velocity profile has its maximum value, $r_{\rm max}$.

We have shown the dependence of the `concentration' terms $r_{\rm
vir}/r_{-2}$ and $r_{\rm vir}/r_{\rm e}$ (and $r_{\rm vir}/R_{\rm e}$)
on the ratio $\rho_{\rm e}/\langle \rho_{\rm univ} \rangle$, where
$\rho_{\rm e}$ and $r_{\rm e}$ (and $R_{\rm e}$) are the Einasto 
(and Prugniel-Simien) scale density and half-mass radius.  We also show how
the slope of these models at $r_{\rm vir}$ depends solely on the ratio
$\rho_{\rm e}/\langle \rho_{\rm univ} \rangle$ (Figure~\ref{Conc}).

Over the radial range $10^{-2} < r/r_{-2} < 4 \times 10^1$, we have shown
both the Einasto and Prugniel-Simien models possess the property that
$\rho(r)/{\sigma(r)}^3$ can be roughly described by a power-law
$r^{-\alpha}$ with the value of $\alpha$ slightly less than 2 for
profile shapes $n$ equal to or greater than the best-fitting values
reported both here and elsewhere. 

Analytical expressions for the logarithmic slope of the Einasto 
and Prugniel-Simien models have been derived, and the slope
expected from the inward extrapolation of these models, inside of
$\sim$0.01 $r_{\rm vir}$, is compared with that from observations of
real galaxies.  The innermost ($r=0$) slope of the Prugniel-Simien
model (0.73--0.87), as currently defined, appears too steep to match
all the galaxy data, but agrees with theoretical expectations for a
slope of $-0.75$ (Taylor \& Navarro 2001) and $-0.78$ (Austin et al.\
2005).
Future work should explore the optimal value of the quantity $p$, the
inner logarithmic profile slope in the Prugniel-Simien model.  Setting
$p=0$, one recovers the Einasto model, which appears capable of
matching the inner profile slopes observed in real galaxies
(Figure~\ref{FigLSB}).  Indeed, the typical value of $\sim -0.7$ at
0.1 kpc in our CDM halos agrees well with the galaxy data from Swaters
et al.\ (2003a) and Simon et al.\ (2005), but is steeper than the
value $-0.2 \pm 0.2$ reported by de Blok et al.\ (2003) and others.
We also note that, at present, the pseudo-isothermal model appears
inconsistent with the galaxy data (Figure~\ref{FigIso}).
% 
% It would be of interest to apply Einasto's model to the
% dark halos of LSB galaxies.

\acknowledgments  

We kindly thank Gary Mamon for his detailed comments on this manuscript. 
We are additionally grateful to Erwin de Blok, Walter Dehnen, and Dean McLaughlin
for their helpful corrections and comments.  We also wish to thank
Peeter Tenjes for kindly faxing us a copy of Einasto's original
papers in Russian. 
A.G.\ acknowledges support from NASA grant HST-AR-09927.01-A
from the Space Telescope Science Institute, 
and the Australian Research Council through Discovery Project Grant DP0451426.
D.M.\ was supported by grants AST 02-06031, AST 04-20920, and AST
04-37519 from the National Science Foundation, and grant NNG04GJ48G
from NASA.
%
% Ben had no support. 
%
J.D.\ is grateful for financial support from the Swiss National
Science Foundation.
B.T.\ acknowledges support from Department of Energy grant G1A62056.


\begin{references}
\reference{Aet94}Abraham, R.G., Valdes, F., Yee, H.K.C., \& van den Bergh, S.\ 1994, ApJ, 432, 75
\reference{aWE06}An, J.H., \& Evans, N.W.\ 2006, ApJ, 642, 752
% \reference{ADK04}Arad, I., Dekel, A., Klypin, A.\ 2004, MNRAS, 353, 15
\reference{Asc04}Ascasibar, Y., Yepes, G., Gottl\"ober, S., \& M\"uller, V, 2004, MNRAS, 352, 1109
\reference{Aet05}Austin, C.G., Williams, L.L.R., Barnes, E.I., Babul, A., \& Dalcanton, J.J.\ 2005, ApJ, 634, 756
\reference{Barne}Barnes, E.I., Williams, L.L.R., Babul, A., \& Dalcanton, J.J.\ 2006, ApJ, 643, 797
\reference{Bla04}Blais-Ouellette, S., Amram, P., Carignan, C., \& Swaters, R.\ 2004, A\&A, 420, 147
\reference{BaN98}Bryan, G., \& Norman, M.\ 1998, ApJ, 321, 80
\reference{Bur95}Burkert, A.\ 1995, ApJ, 447, L25 
\reference{BaO84}Butcher, H., \& Oemler, A.\ 1984, ApJ, 285, 426
% \reference{Car04}Cardone, V.F.\ 2004, A\&A, 415, 839
\reference{CPT05}Cardone, V.F., Piedipalumbo, E., \& Tortora, C.\ 2005, MNRAS, 358, 1325
\reference{Coc04}Coccato, L., Corsini, E.M., Pizzella, A., Morelli, L., Funes, J.G., \& Bertola, F.\ 2004, A\&A, 416, 507
\reference{deB03}de Blok, W.J.G.\ 2003,  Proceedings IAU 220: Dark Matter in Galaxies, Eds.\ S.Ryder, D.J.Pisano, M.Walker, K.Freeman
\reference{deB05}de Blok, W.J.H.\ 2005, ApJ, 634, 227
\reference{dBB02}de Blok, W.J.G., \& Bosma, A.\ 2002, A\&A, 385, 816
\reference{dBM03}de Blok, W.J.G., Bosma, A., \& McGaugh, S.S.\ 2003, MNRAS, 340, 657
\reference{dBM97}de Blok, W.J.G., \& McGaugh, S.S.\ 1997, MNRAS, 290, 533
\reference{dMR01}de Blok, W.J.G., McGaugh, S.S., Rubin, V.C.\ 2001, AJ, 122, 2396
% \reference{Deh93}Dehnen, W.\ 1993, MNRAS, 265, 250
\reference{DMc05}Dehnen, W., \& McLaughlin, D.E.\ 2005, MNRAS, 363, 1057
% \reference{deV48}de Vaucouleurs, G.\ 1948, Ann.\ d'astrophys., 11, 247
\reference{DMS04}Diemand, J., Moore, B., \& Stadel, J.\ 2004, MNRAS, 353, 624 
\reference{Det05}Diemand, J., Zemp, M., Moore, B., Stadel, J., \& Carollo, M.\ 2005, MNRAS, 364, 665
\reference{Ein65}Einasto, J.\ 1965, Trudy Inst.\ Astrofiz.\ Alma-Ata, 5, 87
\reference{Ein68}Einasto, J.\ 1968, Tartu Astr.\ Obs.\ Publ.\ Vol.\ 36, Nr 5-6, 414
\reference{Ein69}Einasto, J.\ 1969, Astrofizika, 5, 137
% \reference{Ei69b}Einasto, J.\ 1969b, Astronomische Nachrichten, 291, 97
\reference{EaH89}Einasto, J., \& Haud, U.\ 1989, A\&A, 223, 89
% \reference{ECF96}Eke, V.R., Cole, S., \& Frenk, C.S.\ 1996, MNRAS, 282, 263
% \reference{FaM00} Ferrarese, L., \& Merritt, D.\ 2000, ApJ, 539, L9
\reference{Fra72}Fraser, C.W.\ 1972, The Observatory, 92, 51
% \reference{Geb00} Gebhardt, K., et al.\ 2000, AJ, 539, L13
\reference{Gent6}Gentile, G., Burkert, A., Salucci, P., Klein, U., \& Walter, F.\ 2006, ApJ, 634, L145
\reference{GMRSZ}Goerdt, T., Moore, B., Read, J.I., Joachim, S/, \& Zemp, M.\ 2006, MNRAS, 368, 1073
\reference{Gra02}Graham, A.W.\ 2002, MNRAS, 334, 721
% \reference{Gra04}Graham, A.W.\ 2004, ApJ, 613, L33
\reference{GaD05}Graham, A.W., \& Driver, S.\ 2005, PASA, 22(2), 118
\reference{Get05}Graham, A.W., Driver, S., Petrosian, V., Conselice, C.J., Bershady, M.A., Crawford, S.M., \& Goto, T.\ 2005, AJ, 130, 1535
\reference{GTC01}Graham, A.W., Trujillo, N., \& Caon, N.\ 2001, AJ, 122, 1707
\reference{HaS05}Hansen, S.H., \& Stadel, J.\ 2006, Journal of Cosmology and Astro-Particle Physics, 5, 14
\reference{Het04}Hayashi, E., et al.\ 2004, MNRAS, 355, 794
% \reference{Hen06}Henriksen, R.N., 2006 (astro-ph/0606654)
\reference{Her90}Hernquist, L.\ 1990, ApJ, 356, 359
% \reference{Jaf83}Jaffe, W.\ 1983, MNRAS, 202, 995
% \reference{JaS02}Jing, Y.P., \& Suto, Y.\ 2002, ApJ, 574, 538
\reference{Krav8}Kravtsov, A.V., Klypin, A.A., Bullock, J.S., \& Primack, J.R.\ 1998, ApJ, 502, 48
\reference{Let06}Lavalle, J., et al.\ 2006, A\&A, 450, 1
\reference{LNGM9}Lima Neto, G.B., Gerbal, D., \& M\'arquez, I.\ 1999, MNRAS, 309, 481
\reference{MMB03}Macci\`o, A.V., Murante, G., \& Bonometto, S.P.\ 2003, ApJ, 588, 35
\reference{MaL5a}Mamon, G.A., \& {\L}okas, E.L.\ 2005, MNRAS, 362, 95
% \reference{MaL5b}Mamon, G.A., \& {\L}okas, E.L.\ 2005b, MNRAS, 363, 705
% \reference{Mam06}Mamon, G.A., \& {\L}okas, E.L., Dekel, A., Stoehr, F., \& Cox, T.J.\ 2006, in the 21st IAP meeting, Mass Profiles and Shapes of Cosmological Structures, ed.\ G.A. Mamon, F.\ Combes, C.\ Deffayet \& B.\ Fort (Paris: EDP) (astro-ph/0601345)
\reference{Mar02}Marchesini, D., D'Onghia, E., Chincarini, G., Firmani, C., Conconi, P., Molinari, E., \& Zacchei, A.\ 2002, ApJ, 575, 801
% \reference{Mar00}M\'arquez, I., Lima Neto, G.B., Capelato, H., Durret, F., \& Gerbal, D.\ 2000, 353, 873
% \reference{Mar01}M\'arquez, I., Lima Neto, G.B., Capelato, H., Durret, F., Lanzoni, B., \& Gerbal, D.\ 2001, A\&A, 379, 767
\reference{MCW06}Mashchenko, S., Couchman, H.M.P., \& Wadsley, J.\ 2006, Nature, in press (astro-ph/0605672)
\reference{Met06}Merritt, D., Graham, A.W., Moore, B., Diemand, J., \& Terzi\'c, B.\ 2006, AJ, submitted (Paper I)
\reference{Met05}Merritt, D., Navarro, J.F., Ludlow, A., \& Jenkins, A.\ 2005, ApJL, 624, L85
% \reference{MaM01} Milosavljevi\'c, M., \& Merritt, D.\ 2001, ApJ, 563, 34
% \reference{Moo98}Moore, B., Governato, F., Quinn, T., Stadel, J., \& Lake, G.\ 1998, ApJ, 499, L5
\reference{Moo99}Moore, B., Quinn, T., Governato, F., Stadel, J., \& Lake, G.\ 1999, MNRAS, 310, 1147
% \reference{MaH03}M\"ucket, J.P., \& Hoeft, M.\ 2003, A\&A, 404, 809
\reference{NFW95}Navarro, J.F., Frenk, C.S., \& White, S.D.M.\ 1995, MNRAS, 275, 720
% \reference{NFW96}Navarro, J.F., Frenk, C.S., \& White, S.D.M.\ 1996, ApJ, 462, 563
% \reference{NFW97}Navarro, J.F., Frenk, C.S., \& White, S.D.M.\ 1997, ApJ, 490, 493
\reference{Net04}Navarro, J.F., et al.\ 2004, MNRAS, 349, 1039
\reference{OKW84}Okamura, S., Kodaira, K., \& Watanabe, M.\ 1984, ApJ, 280, 7
\reference{Pet76}Petrosian, V.\ 1976, ApJ, 209, L1
\reference{Pet05}Prada, F., Klypin, A.A., Simonneau, E., \& Betancort Rijo, J., Santiago, P., Gottl\"ober, S/ \& Sanchez-Conde, M.A.\ 2006, ApJ, 645, 1001
\reference{PaK06}Profumo, S., \& Kamionkowski, M.\ 2006, Journal of Cosmology and Astro-Particle Physics, 3, 3
\reference{PaS97}Prugniel, Ph., \& Simien, F.\ 1997, A\&A, 321, 111
\reference{Ret04}Rasia, E., Tormen, G., \& Moscardini, L.\ 2004, MNRAS, 351, 237
\reference{Rix97}Rix, H.-W., Guhatakhurta, P., Colless, M.M., \& Ing, K.\ 1997, MNRAS, 285, 779
\reference{RaH94}Roberts, M.S., \& Haynes, M.P.\ 1994, ARA\&A, 32, 115
\reference{SaB00}Salucci, P., \& Burkert, A.\ 2000, ApJ, 537, L9
% \reference{SFZ97}Schoenmakers, R.H.M., Franx, M., \& de Zeeuw, P.T.\ 1997, MNRAS, 292, 349
\reference{Ser63}S\'ersic, J.-L.\ 1963, Boletin de la Asociacion Argentina de Astronomia, vol.6, p.41
\reference{Ser68}S\'ersic, J.L.\ 1968, Atlas de galaxias australes
\reference{Sim03}Simon, J.D., Bolatto, A.D., Leroy, A., \& Blitz, L.\ 2003, ApJ, 596, 957
\reference{Sim05}Simon, J.D., Bolatto, A.D., Leroy, A., \& Blitz, L., Gates, E.L.\ 2005, ApJ, 621, 757
\reference{Sota6}Sota, Y.\ Iguchi, O., Morikawa, M., Nakamichi, A.\ 2006, Proceedings of The Third 21COE Symposium : Astrophysics as Interdisciplinary Science (astro-ph/0512587)
\reference{SGH05}Spekkens, K., Giovanelli, R., \& Haynes, M.\ 2005, AJ, 129, 2119
% \reference{Set01}Springel, V., White, S.D.M., Tormen, G., \& Kauffmann, G.\ 2001, MNRAS, 328, 726
\reference{Sto06}Stoehr, F.\ 2006, MNRAS, 365, 147
% \reference{Sto02}Stoehr, F., White, S.D.M., Tormen, G., Springel, V.\ 2002, MNRAS, 335, L84
\reference{Swa3a}Swaters, R.A., Madore, B.F., van den Bosch, F.C., \& Balcells, M.\ 2003a, ApJ, 583, 732
\reference{Swa3b}Swaters, R.A., Verheijen, M.A.W., Bershady, M.A., \& Andersen, D.R.\ 2003b, ApJ, 587, L19
\reference{TaN01}Taylor, J.E., \& Navarro, J.F.\ 2001, ApJ, 563, 483
\reference{THE94}Tenjes, P., Haud, U., \& Einasto, J.\ 1994, A\&A, 286, 753
\reference{TaG05}Terzi\'c, B., \& Graham, A.W.\ 2005, MNRAS, 362, 197
\reference{TGC01}Trujillo, I., Graham, A.W., \& Caon, N.\ 2001, MNRAS, 326, 869
\reference{Vet05}Valenzuela, O., Rhee, G., Klypin, A., Governato, F., Stinson, G., Quinn, T., \& Wadsley, J.\ 2005, ApJ, submitted (astro-ph/0509644) 
\reference{vdB00}van den Bosch, F.C., Robertson, B.E., Dalcanton, J.J., \& de Blok, W.J.G.\ 2000, AJ, 119, 1579
\reference{Ver97}Verheijen, M.A.\ 1997, PhD thesis, Univ.\ of Groningen
\reference{WaK02}Weinberg, M.D., \& Katz, N.\ 2002, ApJ, 580, 627
% \reference{Xet05}Xiao, W., Peng, C., Ye, X., \& Hao, H.\ 2005, ApJL, submitted (astro-ph/0508436)
\reference{Zha96}Zhao, H.S.\ 1996, MNRAS, 278, 488
\end{references}
\end{document}